\documentclass[10pt, reqno]{article}

\usepackage{amsmath,amsfonts,amssymb,amsthm,amsbsy}
\usepackage[dvips]{color}

\theoremstyle{definition}

\newcommand{\abs}[2][]{\ensuremath{#1\lvert#2#1\rvert}}
\newcommand{\bra}[2][]{\ensuremath{#1\langle #2 #1\lvert}}
\newcommand{\com}[3][]{\ensuremath{#1[ #2, #3 #1]}}
\newcommand{\dx}[1][x]{\ensuremath{\text{d}#1}}
\newcommand{\ddx}[2][]{\ensuremath{\frac{\text{d}#1}{\text{d}#2}}}

\newcommand{\ev}[4][]{\ensuremath{#1\langle #2 #1\lvert #3 #1\rvert #4 #1\rangle}}
\newcommand{\ip}[3][]{\ensuremath{#1\langle #2\vert #3 #1\rangle}}
\newcommand{\ket}[2][]{\ensuremath{#1\rvert#2 #1\rangle}}

\newcommand{\pket}[2][]{\ensuremath{#1\rvert #2)}}
\newcommand{\pip}[3][]{\ensuremath{#1( #2, #3 #1)}}
\newcommand{\gbra}[2][]{\ensuremath{{}^{{\scriptscriptstyle {}^{(#1)}}}\!\!\!\!\!\prec #2 \lvert}}
\newcommand{\gket}[2][]{\ensuremath{\rvert #2\!\succ}^{\!\!\!\!{\scriptscriptstyle {}^{(#1)}}}}

\newcommand{\pddx}[2][]{\ensuremath{\frac{\partial #1}{\partial #2}}}

\newcommand{\Tr}{\ensuremath{\operatorname{Tr}}}

\newcommand{\AI}{(\ensuremath{A_{\text{I}}}) }
\newcommand{\AII}{(\ensuremath{A_{\text{II}}}) }
\newcommand{\AIII}{(\ensuremath{A_{\text{III}}}) }
\newcommand{\AIV}{(\ensuremath{A_{\text{IV}}}) }
\newcommand{\AVN}{(\ensuremath{A_{\text{V}}}) }
\newcommand{\AV}{(\ensuremath{A^{\text{old}}_{\text{V}}}) }
\newcommand{\AVP}{(\ensuremath{A^{\text{old}}_{\text{V'}}}) }

\newcommand{\cA}{\mathcal{A}}
\newcommand{\cH}{\mathcal{H}}
\newcommand{\cM}{\mathcal{M}}
\newcommand{\cP}{\mathcal{P}}
\newcommand{\cS}{\mathcal{S}}

\newcommand{\complex}{\mathbb{C}}

\newcommand{\real}{\mathbb{R}}

\newcommand{\ep}{\varepsilon}

\author{A. R. Bohm\thanks{Email:
      \texttt{bohm@physics.utexas.edu}},\ \   Mark Loewe, and Bryan Van de Ven\\\emph{\small{Department of Physics, the
      University of Texas at Austin, Austin, TX 78712-1081, USA}} \\*[0.5cm]}
\title{Time Asymmetric Quantum Theory --- \\\ I Modifying an Axiom of Quantum Physics}

\begin{document}
\maketitle

\begin{abstract}
A slight modification of one axiom of quantum theory changes a
reversible theory into a time asymmetric theory.
Whereas the standard Hilbert space axiom does not distinguish
mathematically between the space of states (in-states of scattering
theory) and the space of observables (out-``states'' of scattering theory)
the new axiom associates states and observables to two different Hardy
subspaces which are dense in the same Hilbert space and analytic in
the lower and upper complex energy plane, respectively.
As a consequence of this new axiom the dynamical equations
(Schr\"{o}dinger or Heisenberg) integrate to a semigroup evolution.
Extending this new Hardy space axiom to a relativistic theory  
provides a relativistic theory of resonance scattering and
decay with Born probablilities that fulfill Einstein causality and the
exponential decay law.
\end{abstract}

\baselineskip=18pt              

\numberwithin{equation}{section}
\setcounter{equation}{0}

\section{Introduction---Time Asymmetry} 
Time asymmetry, irreversibility, time reversal non-invariance are
different concepts and they are (probably) not (all) related to each
other \cite{1.1}, \cite{2a}, \cite{1.2}.
These concepts are usually called arrows-of-time.
Time asymmetry comes from time asymmetric boundary conditions of time
symmetric equations; its most prominent consequence is
causality.
The radiation arrow of time is its example from classical physics.
Irreversibility is usually associated with probability or entropy
increase and called the thermodynamic arrow of time in  classical physics.
In quantum mechanics or quantum statistical mechanics entropy increase is
associated \cite{1.3} with
the effect of the environment or of the measurement apparatus upon the physical 
system.

The possibility of connecting time asymmetry for the solution of the
Maxwell equations to probability was discussed in the Einstein-Ritz
arguments \cite{1.5}, where Einstein maintains that time asymmetric
boundary conditions for the Maxwell equations are not needed and the radiation
arrow of time is a consequence of probability, whereas Ritz insists
that the initial-boundary conditions are the basis of irreversibility.
Peierls \cite{1.6} argued that the implied boundary conditions of
Boltzmann's Stosszahl Ansatz are the origin of irreversibility.

Time-reversal non-invariance is a non-invariance of the dynamical
equations and of the Hamiltonian with respect to the anti-unitary
time-reversal operator \cite{1.4}, and is thus an entirely different
concept and can (probably) not be related to the above arrows-of-time.

In this article we do not want to discuss possible connections between
the arrows of time but 
are only concerned with time asymmetry, i.e. the arrow of time  due to
boundary conditions.

In classical theories one can have time symmetric dynamical equations
with time asymmetric boundary conditions. 

These time asymmetric boundary conditions come in pairs: given one
time asymmetric boundary condition, its time reversed
boundary condition can also be formulated mathematically.
An example is the 
retarded and advanced solutions in classical electrodynamics
(Maxwell equations).
Nature chooses the retarded solutions; 
this is the radiation arrow of time: Radiation must be emitted
first by the source, before it can be detected by the receiver.
Another example is the big bang and big crunch solution of general
relativity (Einstein equations).
The universe expands and the big bang gives us a means of defining the
cosmic time and its origin $t=t_0 =0$.
This is called the cosmological arrow of time.

Quantum physics also has arrows of time.
In terms of experimental arrangements one can formulate it in close
analogy to the radiation arrow of time as the preparation-registration
arrow of time \cite{1.7}: A state must be prepared first by a preparation
apparatus (e.g., accelerator) before an observable can be detected in
it by a registration apparatus (detector).
The radiative decay of an excited state of an atom or a nucleus or
a relativistic particle is the quantum analogue of the classical
radiation arrow of time.

Though the correspondence of the arrows of time for classical
electromagnetic radiation and quantum
radiative decays is obvious, there does not exist such a
correspondence between the respective theories, since standard quantum theory in
Hilbert space (von Neumann(1931)) \cite{1.8} does not allow
time-asymmetric boundary conditions for the Schr\"{o}dinger equation
or the Heisenberg equation.
The Hilbert space axiom, inevitably (by the Stone-von
Neumann theorem, \cite{1.8}, \cite{1.11}) leads to unitary and
therewith reversible time evolution.
In the heuristic scattering theory one circumvents this problem by
using retarded (advanced) Green's functions \cite{1.9} or purely
outgoing boundary conditions \cite{1.10}, which --- very much like the
retarded (advanced) potentials of classical electrodynamics ---
contain time asymmetry.
With the Green's functions one has admitted distributions into the
theory and is outside the Hilbert
space.
It is the conflict between the Hilbert space mathematics and the time
asymmetry of (resonance) scattering described by these
heuristic methods, which lead to such puzzles like violation of
causality or the impossibility of an exponential decay
law.

The axioms of quantum theory are not to be understood as mathematical axioms from which
everything can be derived without using further judgment or
creativity.
An approach of this kind does not appear possible in physics.
The axioms,  or basic assumptions of physics are to be considered as a concise way
of formulating the quintessence of many experimental facts.
As such one could modify them.
One could also leave the old axioms intact and allow their
consequences to partially
disagree with reality;
then the axioms provide an approximate (fuzzy) theory, and since every
theoretical description can only be an 
approximate description, these approximate methods could be adequate
in many respects. 
But if the discrepancy between the consequences of the set of axioms
and the observations becomes too pronounced, one would do well to
make minor changes in the axioms of a theory and devise a modified
theory with wider applicability and greater accuracy.

We want to change just one of the axioms: The Hilbert space axiom  of traditional quantum
mechanics will be modified (infinitesimally) into an axiom using two
(dense) Hardy subspaces of the same Hilbert Space.
The motivation that led to this modification was to obtain a
consistent and exact mathematical theory of resonance scattering and
exponential decay.

\section{Axioms of Quantum Theory}
In quantum mechanics one speaks of states and of observables.
\begin{itemize}
\item[\textbf{\AI}]  
States are mathematically described by state operators (denoted by $\rho$,
$W$) or, if they are pure states, by vectors $\phi$.
Observables are also described by operators $A=(A^\dagger)$,
$\Lambda$, $P=(P^2)$; but if they are projections $P=\ket{\psi}\bra{\psi}$
they can also be described by vectors $\psi$ (up to a phase factor).
The vectors $\psi, \phi \in \Phi$, are elements of a linear space
(pre-Hilbert space) 
  with scalar product $\ip{\psi}{\phi}=\pip{\psi}{\phi}$. 
\end{itemize}
The operators $A, \Lambda \in \cA$, are often assumed to be
elements of an algebra $\cA$ of linear operators in $\Phi$.

In the usual practice of quantum mechanics, 
the space $\Phi$, though often called a Hilbert space, is
treated as a space in which the convergence of infinite sequences is not
a problem, i.e. as a pre-Hilbert space.
One ``kind'' of quantum physical 
systems is associated to a space $\Phi$.
Any vector of $\Phi$ can represent a state or an observable.
The state operator is usually normalized,  $\Tr W = 1$.

Though mathematically one does not distinguish between vectors that
describe states and vectors that describe observables, in experiments,
states and observables are defined by different devices:
\begin{itemize}
\item[\textbf{\AII}] 
A State $W, \phi$ is prepared by a preparation apparatus
  (e.g., accelerator) and
an Observable $\Lambda, \psi$ is registered (or detected) by a
  registration apparatus (detector).
\end{itemize}
Most treatments of the foundations of quantum mechanics agree on
ascribing a \emph{separate} fundamental importance to states and to
observables.
The observed quantities, i.e. the experimental numbers,  are
interpreted as \emph{probabilities to measure an
observable $\Lambda$ in a state $W$} at a time $t$.
\begin{itemize} 
\item[\textbf{\AIII}] 
\text{The probabilities $\cP_W\big(\Lambda(t)\big)$ are calculated in the
theory as Born probabilities}: 
\begin{align}
\cP_{W(t)}\big(\Lambda_0\big)  & = \Tr\big(\Lambda_0W(t)\big)\quad
(\text{Schr\"{o}dinger picture}) \tag{2.1S} \label{eq:2.1S} \\
\cP_{W_0}\big(\Lambda(t)\big)  & = \Tr\big(\Lambda(t)W_0\big)\quad
(\text{Heisenberg picture})\ . \tag{2.1H} \label{eq:2.1H}
\end{align}
\end{itemize}
For the special case that $\Lambda=\ket{\psi^-}\bra{\psi^-}$ and $W =
\ket{\phi^+}\bra{\phi^+}$ this is:
\begin{align}
\cP_{\phi^+}\big(\psi^-(t)\big) & = \abs{\ip{\psi^-}{\phi^+(t)}}^2 \quad
(\text{Schr\"{o}dinger picture}) \tag{2.1'S} \label{eq:1.1'S} \\
 & = \abs{\ip{\psi^-(t)}{\phi^+}}^2\quad
(\text{Heisenberg picture})\ . \tag{2.1'H} \label{eq:1.1'H}
\end{align}
The probabilities $\cP_W(\Lambda)$ are measured as ratios of large
numbers of detector counts
$$
\cP_W\big(\Lambda(t)\big)\big\rvert_{\text{exp}} \approx N_\Lambda(t)/N\ .
$$

\begin{itemize} 
\item[\textbf{\AIV}] 
The time evolution in quantum mechanics is given by the Hamiltonian
operator $H$ and described by the following dynamical equations.

In the Schr\"{o}dinger picture the observables are kept time
independent, and for the state operator $W(t)$, or in the
special case $W = \ket{\phi^+}\bra{\phi^+}$, for the state vector
$\phi^+(t)$ the dynamical equations are
\begin{equation} \tag{2.2S} \label{eq:2S} 
\pddx[W(t)]{t} = \frac{i}{\hbar}\com[\big]{W(t)}{H}
\end{equation} 
or
\begin{equation} \tag{2.3S} \label{eq:3S}
i\hbar \pddx[\phi^+(t)]{t} = H\phi^+(t)\ , \quad \phi^+(t=0)=\phi^+_0\ .
\end{equation}
In the Heisenberg picture the state is kept time independent and for
the observable $\Lambda(t)$, or in the 
special case $\Lambda = \ket{\psi^-}\bra{\psi^-}$, for the observable
vector $\psi^-(t)$  the dynamical equations are
\begin{equation} \tag{2.2H} \label{eq:2H} 
\pddx[\Lambda(t)]{t} = -\frac{i}{\hbar}\com[\big]{\Lambda(t)}{H}
\end{equation} 
or
\begin{equation} \tag{2.3H} \label{eq:3H}
i\hbar \pddx[\psi^-(t)]{t} = -H\psi^-(t)\ , \quad \psi^-(t=0)=\psi^-_0\ .
\end{equation}
\end{itemize}

\addtocounter{equation}{3}

For the sake of simplicity we shall here mainly treat the special case
described by state vectors $\phi^+$ and observable vectors $\psi^-$.
And we have used the notation of scattering theory, $\phi^+$ for the
interaction incorporating (``exact'') in-states and $\psi^-$ for the
interaction incorporating out-observables, since we have the theory
of scattering and decay in mind.

Before we turn to the boundary value conditions imposed on the
solutions of the dynamical differential 
equations (2.3S) or (2.3H)  we want to discuss the calculational methods
used for Born probabilities.

The trace and the scalar product in (2.1) is calculated in the
following way:
\begin{subequations} \label{eq:4}
\begin{equation}  \label{eq:4a}
\Tr(\Lambda W) = \sum_n^N \ev{n}{\Lambda W}{n}
\end{equation}
or
\begin{equation}  \label{eq:4b}
\Tr(\Lambda W) = \int\! \dx[E]\; \ev{E}{\Lambda W}{E}\ .
\end{equation}
For states $W=\ket{\phi}\bra{\phi}$ and observables $\Lambda=
\ket{\psi}\bra{\psi}$ this is given by
\begin{equation}  \label{eq:4c}
\abs{\ip{\psi}{\phi}}^2 = \abs[\bigg]{\sum_n^N \ip{\psi}{n}\ip{n}{\phi}}^2
\end{equation}
or
\begin{equation}  \label{eq:4d}
\abs{\ip{\psi}{\phi}}^2 = \abs[\bigg]{\int\! \dx[E]\;
  \ip{\psi}{E}\ip{E}{\phi}}^2\ .
\end{equation}
\end{subequations}
In order to write and calculate
these formulas one needs the \emph{basis vector expansions}, i.e. the
existence of a complete basis system of eigenvectors $\ket{n}$ or $\ket{E}$.
The basis vector expansions are generalizations of the expansion of a vector $\vec{x}$ in
the 3-dimensional Euclidean space:
\begin{subequations} \label{eq:5}
\begin{equation}  \label{eq:5a}
\vec{x} = \sum_{i=1}^3\vec{e}_i x^i\ .
\end{equation}

The  generalization of \eqref{eq:5a} used in \eqref{eq:4a} and \eqref{eq:4c} is to
an $N$ dimensional (complex) linear scalar product
space.
For $N=\text{finite}$ this eigenvector expansion is well established
for all self-adjoint, normal and unitary operators. 
This means for every $\phi$ in a finite dimensional space $\Phi$ one has
\begin{equation}  \label{eq:5b}
\phi= \sum_{n=1}^{N} \ket{n} \ip{n}{\phi}\ , \qquad N=\text{finite}
\end{equation}
where $\ket{n}$ are eigenvectors of any self-adjoint (or
normal or unitary) operator, usually  
representing a prominent quantum mechanical observable, (e.g., the
Hamiltonian $H$):
\begin{equation}  \label{eq:5c}
H \pket{n} = E_n \pket{n}\ .
\end{equation}
In case $N=\infty$, i.e., for the infinite dimensional (complex)
Hilbert space, not 
all operators important in quantum mechanics have a discrete set of
eigenvalues (discrete spectrum), so that \eqref{eq:5b} with
\eqref{eq:5c} holds only for some observables.
But for all (so far) known quantum systems and all  observables there
are spaces $\Phi$ for which 
a second, continuously infinite generalization of \eqref{eq:5a} holds. 
This means, for every $\phi \in \Phi$ Dirac's continuous
eigenvector expansion holds:.
\begin{equation}  \label{eq:5d}
\phi = \int_{E_0=0}^\infty \! \dx[E]\; \ket{E} \ip{E}{\phi}\ .
\end{equation}
Here $\ket{E}$ is a generalized eigenvector or eigenket
\begin{equation}  \label{eq:5e}
H^\times \ket{E} = E\ket{E}\ , \qquad E_0 \leq E \leq \infty\ .
\end{equation}
This eigenvalue equation precisely means
\begin{equation}  \label{eq:5f}
\ev{\phi}{H^\times}{E} \equiv \ip{H\phi}{E}  = E\ip{\phi}{E} \qquad
\text{for all vectors $\phi \in \Phi \subset \cH$,}
\end{equation}
\end{subequations}
where $\Phi$ is a dense
subspace of $\cH$,
but $\ket{E}$ is \emph{not} an element of $\Phi$ or $\cH$, but $\ket{E}
\in \Phi^\times$, the space of continuous antilinear functionals on $\Phi$.
Dirac had omitted the $\vphantom{H}^\times$ in \eqref{eq:5e}.
$H^\times$ is defined by the first equality \eqref{eq:5f} as a unique extension of
$H^\dagger$, the Hilbert space adjoint operator of $H$, to the space
$\Phi^\times \supset \cH$.

The meaning of the integral in \eqref{eq:5d} and of the eigenkets in
\eqref{eq:5e}, \eqref{eq:5f} depends upon the choice of
the space $\Phi$ of vectors $\{\phi\}$.
Usually one chooses
\begin{subequations}
\begin{equation}  \label{eq:6a}
\phi, \psi \in \Phi\ , \quad\text{where $\Phi$ is the abstract
  Schwartz space.}
\end{equation}
This means that the energy wave function
\begin{equation}\label{eq:6b}
\phi(E) \equiv \ip{E}{\phi} = \overline{\ip{\phi}{E}} \in \cS\ ,
\end{equation}
which corresponds by \eqref{eq:5d} to the vector $\phi$: 
\begin{equation}\label{eq:6c}
\Phi \ni \phi \Leftrightarrow \phi(E) \in \cS\ ,  
\end{equation}
\end{subequations}
is a Schwartz function (infinitely differentiable, rapidly
decreasing  function of $E$).\footnote{The function
  $\phi(E)$ corresponds to the vector $\phi$ in the same way as the
  coordinates $x^i$ corresponds to the vector $\vec{x}$ in
  \eqref{eq:5a}.
One calls this correspondence \eqref{eq:6c} of the abstract linear
topological space $\Phi$ by the function space $\cS$ a ``realization''
of $\Phi$ by $\cS$.}

The spaces $\Phi \subset \cH$ together with the space of antilinear
continuous functionals on $\Phi$ form a Rigged Hilbert Space (RHS) \cite{1}
\begin{subequations} \label{eq:7}
\begin{equation}  \label{eq:7a}
\Phi \subset \cH \subset \Phi^\times
\end{equation}
which is realized by the RHS of Schwartz functions
\begin{equation}\label{eq:7b}
\cS \subset L^2(\real_+, \dx[E]) \subset \cS^\times\ ,
\end{equation}
\end{subequations}
where $\cS^\times$ is the space of tempered distributions ($\real_+$
denotes the positive real semiaxis).
This means that the RHS's \eqref{eq:7a} and \eqref{eq:7b} are
equivalent.\footnote{(there exist bicontinuous operators from
  \eqref{eq:7a} onto \eqref{eq:7b} and vice versa)}
The ordinary Dirac kets are usually defined as $\ket{E}\in \Phi^\times$,
i.e., functionals on the Schwartz space $\Phi$ which fulfill the
eigenvalue relation \eqref{eq:5e} or \eqref{eq:5f}.

The above basic assumptions of quantum mechanics $\AI\cdots\AIV$
 are part of the standard mathematical theory in Hilbert
space \cite{1.8}, including the calculational rules \eqref{eq:4}, except that the kets
 $\ket{E}$ cannot be defined with 
the Hilbert space only.
But using for the integrals in \eqref{eq:4b} and \eqref{eq:4d}
Lebesgue (rather than Riemann) integrals (and admitting for $\phi(E) =
\ip{E}{\phi}$, $\overline{\psi}(E) = \ip{\psi}{E}$ all the elements of
$L^2(\real_+, \dx[E])$ not just the smooth functions) one can also use
\eqref{eq:4b}, \eqref{eq:4d} in Hilbert space quantum mechanics. 

Thus one can add to $\AI\cdots\AIV$ the Hilbert space
axiom: 
\begin{itemize}
\item[\textbf{\AV}]
The set of states $\{\phi^+\}$ is equal to the set of observable
$\{\psi^-\}$ and both are equal to the whole Hilbert space $\cH$:
\begin{equation}\label{eq:8}
\{\phi^+\} \equiv \{\psi^-\} \equiv \cH\ . 
\end{equation}
\end{itemize}
\AI to \AV are the basic assumptions of conventional 
quantum mechanics.

A slightly different version of axiom \AV uses the Rigged Hilbert
space of \eqref{eq:7}.
This version includes the Dirac kets $\ket{E}$ and
the Dirac formalism, i.e., the basis vector expansion \eqref{eq:5d}
and an algebra of observables in $\AI$, which is an algebra of  continuous
(bounded) operators in $\Phi$ (and therefore defined everywhere in
$\Phi$).
This version of \AV states
\begin{itemize}
\item[\AVP]
\begin{equation}\tag{2.8'} \label{eq:8'}
\{\phi^+\} \equiv \{\psi^-\} \equiv \Phi \subset \cH \subset \Phi^\times 
\end{equation}
where $\Phi$ is the Schwartz space.
The Dirac
ket $\ket{E}$ is an element of the space $\Phi^\times$.
\end{itemize}
Using the axioms $\AI\cdots\AIV$ and $\AVP$ one avoids  infinite
energy states and similar pathologies of the 
Hilbert space and one does not require Lebesgue integrals for (2.4b),
(2.4d). 
But one does not describe different physics than with $\AI\cdots \AV$.

An immediate consequence of \AV is that the solutions of the
dynamical equations $\AIV$ with the boundary conditions
$\psi^- \in \cH$, $\phi^+\in\cH$ are for the Schr\"{o}dinger equation
\eqref{eq:3S} given by
\begin{equation}\label{eq:9}
\phi^+(t) = U^\dagger(t)\phi^+_0 = e^{-iHt}\phi^+_0\ , \quad -\infty <
t < \infty
\end{equation}
(or for the general state $W$
\begin{equation} \tag{2.9b}\label{eq:9b}
W(t) = e^{-iHt}W_0 e^{iHt}\ , \quad -\infty < t < \infty\  \text{)}\ ,
\end{equation}
and for the Heisenberg equation \eqref{eq:3H} they are given by
\begin{equation}\label{eq:10}
\psi^-(t) = U(t)\psi^-_0  = e^{iHt}\psi^-_0\ , \quad -\infty <
t < \infty
\end{equation}
(or for the general observable $\Lambda$
\begin{equation} \tag{2.10b}\label{eq:10b}
\Lambda(t) = e^{iHt}\Lambda_0 e^{-iHt}\ , \quad -\infty < t < \infty\  \text{)}\ .
\end{equation}

This is a mathematical consequence of the Hilbert space boundary
condition \AV for the dynamical equations (2.2) and
(2.3) and follows from the Stone-von Neumann theorem
\cite{1.11} for the Hilbert space.
These results mean that the time evolution is given by the unitary group
\eqref{eq:9} or \eqref{eq:10} and is thus time symmetric.
The state $\phi^+$ (in the Schr\"{o}dinger picture) can evolve forward
and backward in time, and the observable $\psi^-$ (in the Heisenberg
picture) can evolve forward and backward and consequently the Born
probabilities (2.1) can be predicted for all positive and
negative values of $t$.
The same follows from the Schwartz space axiom \AVP \cite{12a}.

Quantum theory in Hilbert space is time symmetric.
This is not so bad for the description of spectra and structure of
quantum physical systems, whose states are (or are considered as) stationary.
But this is particularly detrimental for the description of decay
processes and
resonance scattering, which are intrinsically irreversible processes.
There is no consistent theoretical description for decaying states and
resonances in Hilbert space quantum mechanics.
There are only Weisskopf-Wigner methods of which ``there does not
exist...a rigorous theory to which these various methods can be
considered as approximations'' \cite{4}.
Therefore we suggest an alternative hypothesis in place of the Hilbert
space axiom \AV (and $\AVP$).
This new axiom \AVN will be stated using $\AVP$ (the RHS in analogue
of  $\AV$). 
Therefore it will include the Dirac formalism from the start, 
i.e., the kets, the basis vector expansion,
and an algebra of operators without the need to worry about domain questions for operators.
We first shall formulate the axiom \AVN and then will discuss
and motivate it.

\begin{itemize}
\item[\textbf{\AVN}] The set of states defined physically by preparation apparatuses
(accelerator) (e.g., the in-states $\phi^+$ of a scattering experiment)
are mathematically described by
\begin{equation} \label{eq:11}
\{\phi^+\} = \Phi_- \subset \cH \subset \Phi^\times_-\ .
\end{equation}
The set of observables defined by registration apparatuses (detector)
(e.g., the out-observables usually called out-states of a scattering
experiment) are mathematically described by
\begin{equation} \label{eq:12}
\{\psi^-\} = \Phi_+ \subset \cH \subset \Phi^\times_+\ .
\end{equation}
The space $\Phi_-$ and $\Phi_+$ are different (dense) Hardy subspaces of the same Hilbert
space $\cH$.
\end{itemize}

Though observables and states are defined physically as different
entities, the axiom
\AV identifies them mathematically, i.e., $\{\phi^+\} = \{\psi^-\}$.
The same is true for \AVP
(which in scattering theory is called asymptotic completeness).
The new hypothesis $\AVN$ distinguishes \emph{also mathematically} between states
and observables  by assigning them to different dense subspaces of the
Hilbert space $\cH$, the Hardy spaces $\Phi_-$ and $\Phi_+$, respectively.
We shall explain the mathematical properties of Hardy Spaces a little
better in the following section, and first describe here  some
consequences. 

The solutions of the dynamical equation \eqref{eq:3S} with the new boundary
condition \eqref{eq:11} are for the states $\phi^+ \in \Phi_-$, given by
\begin{equation} \label{eq:13}
\phi^+(t) = e^{-iHt}\phi^+\equiv U_-^\dagger(t)\phi^+\ ; \qquad 0 \leq t < \infty\ .
\end{equation}
The solutions of the dynamical equation \eqref{eq:3H} with the new
boundary condition \eqref{eq:12} are
for the observables $\psi^- \in \Phi_+$, given by\footnote{Precisely, the
  semigroup generator $H=H_+$ in \eqref{eq:14} is the restriction
  of the self-adjoint operator $H$ to the (dense in $\cH$) subspace
  $\Phi_+$ and the generator $H=H_-$ in \eqref{eq:13} is the restriction
  of $H$ to $\Phi_-$. 
  The same notation is used for the $U(t)$.
  We often omit the subscripts of the operators
  which are the same as the subscripts of the spaces.}
\begin{equation} \label{eq:14}
\psi^-(t) = e^{iHt}\psi^- \equiv U_+(t)\psi^-\ ; \qquad 0 \leq t < \infty
\end{equation}

Thus, in place of the unitary group solution \eqref{eq:9},
\eqref{eq:10} which one obtains from the dynamical equations (2.2), (2.3) one
obtains under the new boundary conditions \eqref{eq:11}, \eqref{eq:12} the
semigroup solution \eqref{eq:13}, \eqref{eq:14}.

Thus we see that as a consequence of the change of boundary conditions
from \AV to $\AVN$, keeping all other
axioms of quantum mechanics including the dynamical equations the same, we obtain a completely new
situation.
For \AV (and the same for $\AVP$) we obtain the
reversible time evolution \eqref{eq:10}, \eqref{eq:9} given by the unitary
group $U(t) =e^{iHt}$ (or $U^\dagger(t)=e^{-iHt}$) with $-\infty < t <
\infty$.
For \AVN we have only a semigroup time
evolution $0\leq t < \infty$, which cannot be reversed to negative time. 

This singles out a particular time $t_0$, the mathematical semigroup time $t_0
= 0$. 
To interpret this time $t_0$ we calculate the Born probability of the observable
$\ket{\psi^-(t)}\bra{\psi^-(t)}$ in the state $\phi^+$, using the
Heisenberg picture.
From \eqref{eq:14} follows:
\begin{equation} \label{eq:15}
\cP(t) = \abs{\pip{\psi^-(t)}{\phi^+}}^2 =
\abs{\pip{e^{iH_+t}\psi^-}{\phi^+}}^2\ , \quad \text{ for $t \geq 0$ only.}
\end{equation}
The same result one obtains using the Schr\"{o}dinger picture for the
probability of the observable $\ket{\psi^-}\bra{\psi^-}$ in the state
$\phi^+(t)$.
The prediction $t\geq 0 =t_0$ means that the probability for an
observable $\psi^-(t)$ in a state $\phi^+$ makes sense only for times
$t \geq t_0=0$.
This is a mathematical consequence of the new hypothesis
$\AVN$.

Whereas a group like \eqref{eq:9}, \eqref{eq:10} does not have a
distinguished time $t_0$ since $-\infty < t< \infty$, the semigroups
\eqref{eq:13} and \eqref{eq:14} introduce a distinguished time $t_0=0$.
We interpret this $t_0$ as the time at which the state has been prepared and
at which the registration of an observable in this state can start.
The existence of such a time $t_0$ in quantum mechanics is fairly obvious, because, 
as stated in Section 1, a state must be prepared first (by $t_0$) before an observable
can be detected in it. 
The detailed interpretation of $t_0$ and its
determinations in each particular process can, however, be quite
intricate as will be discussed below and in a subsequent paper \cite{5}.

The conclusion of this section is that we can have two systems of
axioms.\footnote{The axioms $\AI\cdots\AIV, \AV$ do not constitute the
  complete system of axioms for quantum mechanics, we have mentioned 
  here only those axioms that are relevant for the discussions of this paper.}
They differ from each other in the one axiom that specifies the
boundary conditions for the solutions of the dynamical equations.
All other axioms agree with each other and are the same as formulated
or practiced in the traditional quantum mechanics.
Using the set of axioms $\AI, \ldots, \AIV$ and
\AV one has the conventional, time symmetric
quantum mechanics with reversible time evolution, using the set of
axioms 
$\AI, \ldots, \AIV$ and \AVN
one obtains a time asymmetric quantum theory.
It is in particular the choice of the Hardy spaces for the axiom \AVN
that leads to 
time asymmetry.

\section{Similarities and Differences with the Traditional Practices}

After having decided that the states $\phi^+$ and the observables
$\psi^-$ have their own spaces, $\phi^+ \in \Phi_-$ and $\psi^- \in
\Phi_+$ respectively,\footnote{The notation ${}^+$ and
  ${}^-$ for the vectors describing the in-states and
  out-observables is nearly standard in the physicists scattering
  theory \cite{1.9}.
  We therefore want to use it here for the vectors $\phi^+$
  (in-states) and $\psi^-$ (out-observables, often called out-states,
  cf. \cite{5} section 3) and their basis vectors $\ket{E, b^\pm}$,
  the plane wave solutions of the Lippmann-Schwinger equation \eqref{eq:4.6}.
  The labels ${}_-$ and ${}_+$ of the
  spaces refer to the standard notation that mathematicians use for Hardy
  spaces of the lower and upper complex semiplane.
  Since we want to retain the physicists and the mathematicians
  conventions we have to accept the mismatch in the notation of
  \eqref{eq:11} and \eqref{eq:12}, where the physics of prepared
  in-states and detected out-observables is mapped to the Hardy spaces
  of the lower and upper complex energy plane, respectively.} 
each of them must have their own linear basis vector expansion.
We denote their respective basis kets as $\ket{E, b^+} \in
\Phi^\times_-$ and $\ket{E, b^-} \in \Phi^\times_+$, where the $\ket{E,
  b^\pm}$ are generalized (in the sense of \eqref{eq:5f}) eigenvectors
of the exact Hamiltonian $H=H_0+V$, 
\begin{equation} \label{eq:4.1}
H\ket{E, b^\pm} = E \ket{E, b^\pm}\ , \quad E_0 \leq E < \infty
\end{equation}
the $b$ in \eqref{eq:4.1} are degeneracy labels (e.g., the angular
momentum quantum numbers $j, j_3$).
The basis vector expansions of the in-states $\phi^+ \in \Phi_-$ and
the out-observables $\phi^- \in \Phi_+$ are given by:
\begin{align}
\Phi_+ \ni \psi^- &= \sum_b \int_0^\infty \!\dx[E]\; \ket{E,b^-}
\ip{{}^-\!E, b}{\psi^-}\ , \label{eq:4.2}\\
\Phi_- \ni \phi^+ &= \sum_b \int_0^\infty \!\dx[E]\; \ket{E,b^+}
\ip{{}^+\!E, b}{\phi^+}\ . \label{eq:4.3}
\end{align}
These are analogous to the Dirac basis vector expansion (2.5d) and
mathematically they are the Nuclear Spectral Theorem \cite{1} for the
two RHS's \eqref{eq:11}, \eqref{eq:12}.
The energy wave functions $\ip{{}^+E}{\phi^+}$ and
$\ip{{}^-E}{\psi^-}$ describe the apparatus that prepares the
state 
$\phi^+$ and the apparatus that detects the observable $\psi^-$, respectively:
\begin{align}
\phi^+(E) & \equiv \ip{{}^+E}{\phi^+} = \ip {E}{\phi^{\text{in}}} \quad
\text{is given by the energy distribution of the prepared incident
  beam,} \label{eq:4.4} \\
\psi^-(E) & \equiv \ip{{}^-E}{\psi^-} = \ip {E}{\psi^{\text{out}}} \quad
\parbox[t]{4in}{is given by the energy resolution of the
  detector and measures the energy distribution of the detected
  observables (out-states).}\label{eq:4.5}
\end{align}

We have thus two sets of basis vectors $\ket{E,b^\pm}=\ket{E^\pm} \in
\Phi^\times_\mp$ corresponding to the two RHS \eqref{eq:11} and
\eqref{eq:12}.
The analogy of \eqref{eq:4.2} and \eqref{eq:4.3} with the
conventional scattering theory suggests that the $\ket{E,b^\pm}$
correspond to the in- and out- plane wave ``states'' $\ket{E^+}$ and
$\ket{E^-}$.
These plane wave states are conventionally specified as solutions of the
Lippmann-Schwinger equations, which are used to describe a pair of time
asymmetric boundary conditions in a heuristic way:
\begin{equation} \label{eq:4.6}
\ket{E^\pm} = \ket{E} + \frac{1}{E - H\pm i\ep} V\ket{E} = \Omega^\pm
\ket{E}\ ,
\end{equation}
where $(H-V)\ket{E} = E\ket{E}$.

We shall therefore also call our $\ket{E,b^\pm}$, which are
mathematically defined as the functionals on the Hardy spaces
$\Phi_\mp$, Lippmann-Schwinger kets.
They are more general than the ordinary Dirac kets which are defined as
functionals on Schwartz spaces.

The Schwartz energy wave functions $\ip{E}{\phi} \in S$ are smooth
(infinitely differentiable), rapidly decreasing functions on the real
positive energy axis.
From the $\pm i\ep$ in energy, $\ket{E^\pm} = \lim\limits_{\ep\to+0}\ket{E\pm i\ep^\pm}$, of the
Lippmann-Schwinger equation \eqref{eq:4.6} we can conclude that the
energy wave functions 
\begin{equation} \label{eq:4.7}
\phi^+(E) = \ip{{}^+\!E}{\phi^+} = \overline{\ip{\phi^+}{E^+}}
\end{equation}
must be analytic in the lower half plane, and the energy wave
functions
\begin{equation} \label{eq:4.8}
\psi^-(E) = \ip{{}^-\!E}{\psi^-} = \overline{\ip{\psi^-}{E^-}}
\end{equation}
must be analytic in the upper half plane, at least in an infinitesimal
strip below and above the real axis, respectively.

From this we conjecture that the energy wave functions \eqref{eq:4.4}
and \eqref{eq:4.5} should not only be smooth functions on the real
axis but they should only be those smooth functions that can be
analytically continued into the lower (for \eqref{eq:4.7}) and upper
(for \eqref{eq:4.8}) complex energy semiplanes.
We also would want them to vanish rapidly enough when one goes to the
infinite semicircle.
This is essentially the definition of Hardy functions on the lower and
upper semiplane (for a definition see \cite{6}).
To make this into a precise hypothesis we \emph{postulate}:\footnote{A
  consequence of \eqref{eq:4.10} is that $\ip{\psi^-}{E-i\ep^-} \in
  \cH^2_- \cap S\|_{\real_+}$ and similar for \eqref{eq:4.9}.}
\begin{align}
\ip{{}^+\!E-i\ep}{\phi^+} = \overline{\ip{\phi^+}{E+i\ep^+}} \in
\cH^2_-\cap S \|_{\real_+}\ , \label{eq:4.9} \\
\ip{{}^-\!E+i\ep}{\psi^-} = \overline{\ip{\psi^-}{E-i\ep^-}} \in
\cH^2_+\cap S \|_{\real_+}\ , \label{eq:4.10}
\end{align}
where $\cH^2_\mp \cap S\|_{\real_+}$ is the space of smooth ($\in S$)
Hardy functions on the lower/upper complex $E$-plane on the \emph{second
sheet} of the Riemann surface for the $S$-matrix.
That we choose the second sheet of the $S$-matrix is related to
the analyticity property of the $S$-matrix.
This postulate \eqref{eq:4.9}, \eqref{eq:4.10} is the new axiom \AVN because $\Phi_\mp$
are defined as the abstract linear topological spaces which are
realized by the function spaces $\cH^2_\mp \cap S\|_{\real_+}$.
This means, in analogy to \eqref{eq:7a}, \eqref{eq:7b}, that the
following 
triplets are equivalent:
\begin{equation} \label{eq:4.11}
\phi^\pm \in\Phi_\mp \subset \cH \subset \Phi_\mp^\times
\quad\Longleftrightarrow\quad \ip{{}^\pm\!E}{\phi^\pm} \in \cH^2_\mp \cap
S\|_{\real_+} \subset L^2(\real_+) \subset ( \cH^2_\mp \cap
S\|_{\real_+})^\times\ .
\end{equation}
The important property of the Hardy space triplets is that they are
indeed Rigged Hilbert spaces \cite{7}, so that the basis vector
expansions \eqref{eq:4.2} and \eqref{eq:4.3} are fulfilled as the
nuclear spectral theorem for the RHS's in \eqref{eq:4.11}. 

With the arguments that led from \eqref{eq:4.6} to \eqref{eq:4.9}, \eqref{eq:4.10} we have
given a heuristic justification of the Hardy space axiom $\AVN$.
A more compelling argument (which exceeds the scope of this paper)  is
that only with the Hardy space 
properties can one obtain a theory that relates Breit-Wigner
resonances to exponential decay \cite{8}, \cite{9}.

From \AVN follows the time asymmetry \eqref{eq:13} \eqref{eq:14} and
\eqref{eq:15}. \cite[sect. 5.6]{8}
The important mathematical theorem behind this time asymmetry 
is the Paley-Weiner Theorem \cite[Appendix A]{8} for Hardy functions.
The choice
of the Schwartz space $S$ in \eqref{eq:4.11} is not crucial for time asymmetry.
Therefore  $S$ could be---and will be in the relativistic case---replaced
by some other suitable spaces.

With the Hardy Rigged Hilbert Spaces \eqref{eq:4.11}  we have a wealth of new
mathematical objects which are not contained in $\cH$ or in the
Schwartz RHS \eqref{eq:7}.
In addition to apparatus prepared states $\phi^+ \in \Phi_-$ with
smooth, analytic in $\complex_-$, wave function $\phi^+(E) =
\ip{{}^+\!E}{\phi^+}$, describing the energy distribution
$\abs{\phi^+(E)}^2$ of accelerator beam, and in addition to the
detected observables $\psi^- \in \Phi_+$ with smooth, analytic in
$\complex_+$ wave functions  $\psi^-(E) = \ip{{}^-\!E}{\psi^-}$ describing the
energy resolution of the detector one has in RHS's \eqref{eq:11} and
\eqref{eq:12}, generalized vectors (continuous antilinear functionals on $\Phi_\pm$) or kets
$F^\mp \in \Phi_\pm^\times$.

An example of these kets are the Lippmann-Schwinger scattering states
which are the generalized eigenvectors \eqref{eq:4.1} of the exact Hamiltonian $H$ with
real eigenvalues given by the scattering energies.
In addition to these eigenkets with real eigenvalues\footnote{They are
  boundary values of kets with complex eigenvalues in the upper and
  lower complex half plane, respectively.},
there are many more functionals $F^\mp \in \Phi_\pm^\times$.
In particular, there are generalized eigenvalues of the self-adjoint
Hamiltonian $H$ with complex eigenvalue $z \in \complex_\pm$.
One special example of these are the exponentially decaying state
vectors $\psi^G = \ket{E_R - i\Gamma/2, b^-}\sqrt{2 \pi \Gamma} \in
\Phi_+^\times$ which are eigenkets of the self-adjoint Hamiltonian $H$
with complex eigenvalue,
\begin{equation} \label{eq:4.12}
H^\times \psi^G = (E_R-i\Gamma/2) \psi^G\ ,
\end{equation}
where the generalized eigenvalue $E_R-i\Gamma/2$ is the position of
the resonance pole of the $S$-matrix in the complex energy plane, second
Riemann sheet.
These generalized eigenvectors we call Gamow vectors or Gamow kets.
They describe exponentially decaying resonance states given by first
order poles of the $S$-matrix; their generalization to relativistic
physics is the subject of the two subsequent papers \cite{5} \cite{10}.

There are many more special vectors in $\Phi^\times_+$ (and in
$\Phi_-^\times$) than the Lippmann-Schwinger kets \eqref{eq:4.1} and
the Gamow kets \eqref{eq:4.12}.
Some of these, the Gamow-Jordan kets associated to
higher than first order poles of the $S$-matrix,
will be presented in the Appendix \cite{11} \cite{12}.
The states associated to higher order $S$-matrix poles have the
following curious features: 1.) they have---contrary to the
standard opinion \cite{ref27}, \cite{new32}---an exponential time
evolution \cite{12}, and 2.) they are not described by state vectors
but by density operators that cannot be reduced further into ``pure''
states given by dyadic products \cite{12}.
Other examples of decaying mixed states are associated to cuts in the
lower half-plane, their asymmetric time evolution is described by the
extended Liouvillian \cite{new33}.
We want to restrict our discussions in this introductory paper on time
asymmetric quantum theory to the simplest and experimentally
well-documented case of a first order pole resonance and to the Gamow
states \eqref{eq:4.12}.

Using the bra-ket (i.e., the functional $ F^-(\psi^-)=\ip{\psi^-}{F^-}$ for all
$\psi^- \in \Phi_+$) the time evolution of the generalized vector $F^-(t) \in \Phi_+^\times$ can be
defined as 
\begin{equation} \label{eq:4.13}
\begin{split}
\ip{\psi^-(t)}{F^-} & = \ip{e^{iHt}\psi^-}{F^-} \\
  & = \ip{\psi^-}{e^{-iH^\times t}F^-} \\
  & \equiv \ip{\psi^-}{F^-(t)} \qquad \text{for $t\geq 0$ only}
\end{split}
\end{equation}
since \eqref{eq:14} (and also \eqref{eq:13}) holds only for $t \geq 0$.\footnote{Precisely, the
  semigroup generator $H=H_+$ in \eqref{eq:14} is the restriction of
  the self-adjoint operator $H$ to the (dense in $\cH$) subspace
  $\Phi_+$ and the generator $H=H_-$ in \eqref{eq:13} is the
  restriction of $H$ to $\Phi_-$. 
  For simplicity of notation we have omitted
  the subscripts for the operators.
  But in order to make \eqref{eq:4.13} precise, we have to be more
  accurate in our notation and specify the space on which the
  operators act.
  The semigroup $(e^{iH_+t})^\times$ acts on $\Phi_+^\times$ and has the
  generator $H_+^\times$, $(e^{iH_+ t})^\times = e^{-iH_+^\times t}$.
  Analogously, the semigroup $(e^{-iH_-t})^\times = e^{iH^\times_-t}$ acts on $\Phi_-^\times$
  and has the generator $H_-^\times$.
  $H_+^\times$ is a uniquely defined extension of $H_+$ and of
  $H=H^\dagger$. 
  The operators $H, H^\dagger$ are generators of the unitary group
  \eqref{eq:9} and \eqref{eq:10} in $\cH$ (where the differentiation $H = \frac1{i}
  \ddx[U(t)]{t}\Big|_{t=0}$ is defined with respect to the topology
  $\cH$.)
  The operators $H_+$ and $H_+^\times$ are generators of the semigroup
  \eqref{eq:14} and of the semigroup \eqref{eq:4.15}, respectively.
  This means that the differentiation $H_+ =\frac{1}{i} \ddx[U_+]{t}$
  and $H_+^\dagger = -\frac{1}{i} \ddx[U_+^\dagger]{t}$ is defined with respect to the
  topology in $\Phi_+$ and $\Phi_+^\times$ respectively.
  That the restrictions and extensions of the generators of the group
  \eqref{eq:10} are the generators of the semigroup in $\Phi_+$ and of
  its
  conjugate in $\Phi_+^\times$ is highly non-trivial.}
$F^-(t)$ is again a solution of the Schr\"{o}dinger equation
\begin{equation} \label{eq:4.14}
i\pddx{t} F^-(t) = H^\times F^-(t) \quad\text{but with boundary
  condition} \quad F^-(0) = F^-\in \Phi_+^\times\ .
\end{equation}
Thus $F^-(t)$ represents a generalized state with semigroup time
evolution
\begin{equation} \label{eq:4.15}
F^-(t) = U^\times(t) F^- = e^{-iH^\times t}F^- \quad \text{for } t\geq
0\ .
\end{equation}

In analogy to the Born probabilities we would want to interpret the
matrix elements \eqref{eq:4.13} as something like probability
amplitudes  (as done for Dirac ket, where $\abs{\ip{\psi}{E}}^2$ is
the probability density for $E$).
This is a generalization of (2.1').

The probability to measure an observable
$\ket{\psi^-(t)}\bra{\psi^-(t)}$ (or $\Lambda^-(t)$) in the
generalized state $F^- \in \Phi^\times_+$ is given by
\begin{equation} \label{eq:4.16}
\cP_{F^-}(t) =\abs{\ip{\psi^-(t)}{F^-}}^2 = \abs{\ip{\psi^-}{F^-(t)}}^2
\end{equation}
and it is defined for $t\geq 0$ only.

As mentioned in Section 1, 
the semigroup time $t=0$ will be interpreted as  any arbitrary but finite time $t_0$
at which the generalized state
described by $F^-$ has been prepared.
The quantum system represented by the generalized state vector
$F^-$ is an ensemble of individual micro-particles prepared or created
under identical conditions.
The generalized state $F^-$ starts its dynamical evolution
generated by the Hamiltonian $H^\times \supset H$ at the semigroup
time $t=0$.
An experiment in quantum physics is done on an ensemble of
individual micro systems prepared under identical conditions; this
does  not mean that they are prepared at the same time $t_0$ in the life of the experimentalist.
In most cases, the semigroup time $t=0$ means, in fact, a collection of times $t_0^{(1)}$,
$t_0^{(2)}, \ldots, t_0^{(i)}, \ldots$ and how these times are
measured will be discussed in the subsequent paper \cite{5}.
All these times $t_0^{(i)}$ are represented by the mathematical
semigroup time $t=0$ of the state $F^-(t)$ that describes the
ensemble of microsystems.

If we choose for $F^-$ the Gamow vector $\psi^G$ we obtain from
\eqref{eq:4.12} 
\begin{equation} \label{eq:4.17}
\psi^G(t) = e^{-iH^\times t} \psi^G = e^{-iE_R t}e^{-\frac\Gamma{2}
  t}\psi^G\ ;\qquad t\geq 0\ .
\end{equation}
For the probability rate to detect the decay products $\psi^- \in
\Phi_+$ in state $\psi^G \in \Phi_+^\times$ at time $t$  we obtain then
\begin{equation} \label{eq:4.18}
\cP_{\psi^G(t)} =\abs{\ip{\psi^-(t)}{\psi^G}}^2 = e^{-\Gamma
  t}\abs{\ip{\psi^-}{\psi^G}}^2 \qquad \text{for $ t\geq 0$ only.}
\end{equation}

The time asymmetry $t\geq 0$ means that no probability is predicted
for times $t$ before the quantum system was prepared at $t=0$, as is
in agreement with all observations (causality).
Similar arguments also apply to the Lippmann-Schwinger kets $F^- = \ket{E^-}
\in \Phi_+^\times$, for which this time asymmetry has been
unrecognized: 
\begin{equation} \label{eq:4.19}
\ip{\psi^-(t)}{E^-} = \ip{e^{iHt}\psi^-}{E^-} =
  \ev{\psi^-}{e^{-iH^\times t}}{E^-} = e^{-i E t}
  \ip{\psi^-}{E^-}\qquad  \text{for $ t\geq 0$ only} 
\end{equation}
or as a functional equation in the space $\Phi_+^\times$:
\begin{equation} \label{eq:4.20}
e^{-iH^\times t} \ket{E^-} = e^{-iEt}\ket{E^-} \qquad \text{for $t\geq 0$ only.}
\end{equation}
This time asymmetry is the difference between the Lippmann-Schwinger kets and the
ordinary Dirac kets, defined as functionals on the Schwartz space
$\ket{E} \in \Phi^\times$.  
The Schwartz space kets fulfill:
\begin{equation} \label{eq:4.21}
e^{-iH^\times t}\ket{E} = e^{-iEt}\ket{E} \qquad \text{for all }
-\infty < t < \infty\ ,
\end{equation}
precisely
\begin{equation}
\ip{e^{iHt\psi}}{E} = e^{-iEt}\ip{\psi}{E} \quad \text{for all $\psi
    \in \Phi$}.
\end{equation}
And this has always been assumed for the Dirac kets, even when they
were not precisely defined as functionals.
On the dual of the Schwartz space $\Phi^\times$ the extension
$H^\times \supset H^\dagger = H$ of the self-adjoint $H$ generates a group, on
the dual of the Hardy space $\Phi_+^\times$ the extension
$H_+^\times \supset H^\dagger = H$ of the self-adjoint $H$ generates a semigroup.

It is important to realize that the popular Lippmann-Schwinger kets
$\ket{E^\pm}$ cannot fulfill the following two conditions
simultaneously:
\begin{itemize}
\item[1)] Be boundary values of ``analytic kets'' in the complex
  half-plane,\footnote{This means that the functions
  $\ip{{}^+\!E}{\phi^+}$ and $\overline{\ip{{}^-\!E}{\psi^-}} =
  \ip{\psi^-}{E^-}$ are analytic functions in the lower complex
  halfplane ($\ip{{}^-\!E}{\psi^-}$ analytic in the upper half-plane).}
  as indictated by the $i\ep$ in \eqref{eq:4.6}.
  \item[2)] Fulfill the unitary group evolution \eqref{eq:4.21}.
\end{itemize}
One of these conditions has to go.
For the ordinary Dirac kets (functionals on the Schwartz space) one
keeps \eqref{eq:4.21}; then one cannot have the analyticity required
for the in- or out- boundary condition.
For the Lippmann-Schwinger kets one keeps the time asymmetric boundary conditions,
because that was the purpose for introducing them; then they cannot fulfill
\eqref{eq:4.21}. 
But, --- after turning them into mathematically well defined objects
by specifying the spaces $\Phi_\mp$ on which they are
eigenfunctionals---they fulfill \eqref{eq:4.19}.

\section{Conclusion}
We have modified the system of traditional axioms of quantum mechanics
slightly by exchanging one of its axioms, the Hilbert space axiom
$\AV$, for the Hardy space axiom $\AVN$. 
This new axion \AVN distinguishes between observables
and states and describes them by two different (dense) subspaces of
the same Hilbert space.
The idea of using two different spaces for two kinds of ``states'' has
been mentioned before in footnote 14 of the historical paper \cite{18a}.
Feynman distinguishes between the ``state at times $t' < t_0$ defined
by the preparation'' (our prepared states $\{\phi^+\}$) and the
``state characteristic of the experiment''at times $t'' > t_0$ (our
detected observables $\{\psi^-\}$).
The possibility, that $\phi^+$ and $\psi^-$ are from two different
spaces, he mentions in footnote 14 attributing it to H. Snyder, but
does not consider it any further.
Here we have implemented this possibility by choosing for $\{\phi^+\}$
and $\{\psi^-\}$ the two different Hardy spaces $\Phi_-$ and $\Phi_+$
which are related by the Paley-Wiener theorem \cite{6} to $t' < t_0 =0$
and $t'' > t_0 =0$, respectively.

The new axiom  $\AVN$,  $\{\phi^+\}
\subset \Phi_-$, means that the energy distribution in the accelerator beam
$\abs{\phi^+(E)}^2$ is described by a smooth rapidly decreasing
function $\phi^+(E)$ that, additionally, can be analytically
continued into the \emph{lower} half complex energy plane.
Similarly, 
$\{\psi^-\} \subset \Phi_+$ means that the energy resolution
of the detector $\abs{\psi^-(E)}^2$ is described by a smooth rapidly decreasing
function $\psi^-(E)$ that can be analytically continued into the \emph{upper}
half complex energy plane.

In contrast, the  Hilbert space axiom $\AV$ states that
$\phi^+, \psi^- \in \cH$, which means  
that  the energy distributions  $\abs{\psi^-(E)}^2$ and
$\abs{\phi^+(E)}^2$  are both described by Lebesgue 
square-integrable functions $\psi^-(E)$ and $\phi^+(E)$.
The modified version \AVP of this axiom (which is just a refinement of
\AV justifying many of the calculational tools that physicists use)
means that the energy distribution functions 
$\psi^-(E)$ and $\phi^+(E)$ can be given by any smooth rapidly decreasing
(Schwartz space) functions.
In both cases, $\AV$ and $\AVP$, it does not matter whether the function can
be analytically continued into the complex energy plane or not.
Thus the Hardy space hypothesis differs from the old axiom \AVP only by the additional
requirement that the wave function of the states $\phi^+(E)$ can be
analytically continued into the lower complex energy half-plane and the wave
functions of the
observables $\psi^-(E)$ can be analytically continued into the upper
complex energy
half plane.

Observing whether or not an energy wave function can be analytically
continued to complex energies using the energy resolution of the apparatus
does not appear possible.
Thus the two axioms \AVP and \AVN can not be distinguished from
each other by direct observations.

However, the differences in the consequences of \AVP and \AVN are enormous.
It is remarkable that such minor, 
and experimentally imperceptible changes in the mathematics (topology)
of the axioms can lead to
such enormous changes in the consequences of the mathematical theory.

The consequences of the Hardy space axiom
\AVN are: 
\begin{enumerate}
\item a consistent mathematical theory of resonance scattering and decay for
which the lifetime-width relation $\tau = \hbar/\Gamma$ is an exact
property of the Gamow state which is the new idealized physical notion
provided by the Hardy space \cite{8}, and 
\item time asymmetry and causality versus time symmetry and problems
  with causality for $\AV$.
\end{enumerate} 
The dominant opinion about quantum mechanics---supported by the Axiom $\AV$---
is that the time evolution (for isolated systems) 
is reversible.
It is described by the unitary group \eqref{eq:9} with the reverse
evolution of $U(t)$ given by  $U^{-1}(t) = U(-t)$.
But the idea of time asymmetry in quantum mechanics has a long history
and is connected with many distinguished names.
It probably goes back to R. Peierls and his school (1938), who formulated it
in terms of purely outgoing boundary conditions \cite{1.10} \cite{23n}
for the Schr\"{o}dinger equation \eqref{eq:3S}.
The irreversible nature of quantum decay has been mentioned in
\cite{24n} and in a few monographs \cite{2a} \cite{25n}; T. D. Lee called it the
``impossibility of constructing time reversed quantum solutions for a
microphysical system.''
That the irreversibility should be intrinsic, rather that caused by the
external effects of a quantum reservoir or the environment was
emphasized by Prigogine and his school \cite{26n}.
Gell-Mann and Hartle \cite{27n} call it a fundamental arrow of time
and refer to Feynman \cite{18a} when they
use time asymmetry in order to
avoid inconsistencies for the probabilities of histories in their
quantum mechanics of cosmology. 

The axioms of non-relativistic time-asymmetric quantum mechanics
can be extended to a relativistic theory by an appropriate
extension of the Hardy space axiom into the relativistic domain.
There the hypothesis $\AVN$ will lead to new
predictions.
This relativistic theory of resonance scattering and decay, in which
the time evolution semigroup is generalized to the causal Poincar\'{e}
semigroup, is the subject of subsequent papers \cite{5}, \cite{10}.

\bigskip

\noindent\textbf{Acknowledgement} \\
We gratefully acknowledge helpful discussions on the subject of this
paper with M. Gadella and S Wickramasekara and valuable support from
the Welch Foundation.
We are also thankful to W. Drechsler for discussions on the subject of
this paper and for pointing out the statement in \cite{18a}.

 \begin{appendix}

\section*{Appendix A Gamow Jordan Vectors}
Though the operator $H^\dagger$ in Hilbert space is a self-adjoint
operator its (unique) extension $H^\times \supset H^\dagger$ in the
space $\Phi_+^\times \supset\cH$ is not.
Self-adjoint operators in $\cH$ and in finite dimensional subspaces
thereof are always diagonalizable.
But $H^\times$ is not self-adjoint or normal on all finite
dimensional subspaces of $\Phi^\times_+$.
Therefore there may be subspaces on which $H^\times$ is not
diagonalizable.
This does not happen if the $S$-matrix has only first order pole
singularities, but it will happen if the $S$-matrix has a second or in
general $r$th (finite) order pole \cite{11}, \cite{12}.
In the
same way as one derives the Gamow vectors \eqref{eq:4.12} from the
first order pole, one derives 
from the $r$th order $S$-matrix pole at $z_r = E_r - i \Gamma/2$ on
the second (or higher) Riemann sheet $r$ Gamow vectors of order $k=0,1,2,\ldots, (r-1):$ 
\begin{equation}\tag{A1} \label{eq:a1}
\gket[0]{z_R^-},\gket[1]{z_r^-}, \ldots, \gket[r-1]{z_r^-}\ . 
\end{equation}
The $k$th order Gamow vectors $\gket[k]{z_r}$, $k=0,1,,2,
\ldots, r-1$ are Jordan vectors of degree $k+1$. \cite{13}
They fulfill the ``generalized'' eigenvector equation 
\begin{equation}\tag{A2} \label{eq:a2}
(H^\times-z_R)^{k+1}\gket[k]{z_R^-}=0\ .
\end{equation}
and fulfill in detail
\begin{equation}\tag{A3} \label{eq:a3}
\begin{split} 
H^\times\gket[0]{z_R^-} & = z_R \gket[0]{z_R^-} \\
H^\times\gket[0]{z_R^-} & = z_R \gket[1]{z_R^-}+ \Gamma\gket[0]{z_R^-}  \\
 & \qquad \vdots  \\
H^\times\gket[k]{z_R^-} & = z_R \gket[k]{z_R^-}+ \Gamma\gket[k-1]{z_R^-} \\
 & \qquad \vdots  \\
H^\times\gket[r-1]{z_R^-} & = z_R \gket[r-1]{z_R^-}+
\Gamma\gket[r-2]{z_R^-}\ .
\end{split}
\end{equation}
This means $\gket[k]{z_r^-} \in \Phi_+^\times$ and the $r$th order
$S$-matrix pole is associated to an $r$ dimensional subspace $M_{z_r}
\subset \Phi_+^\times$, spanned by the $\gket[k]{z_r^-}$,
$k=0,1,2,\ldots, (r-1)$, i.e., to the set of all $F^-_{z_r} \in M_{z_r}
\subset \Phi_+^\times$ is 
\begin{equation}\tag{A4} \label{eq:a4}
F^-_{z_r} = \sum_{k=0}^{r-1} \gket[k]{z_r^-} c_k\ , \qquad c_k
\in \complex\ .
\end{equation}
On $M_{z_r} \subset \Phi_+^\times$ the Hamiltonian $H^\times$ (i.e.,
the extension of the self-adjoint operator $H^\dagger$ to
$\Phi^\times$) is not diagonalizable, but can only be brought into the
normal form of a Jordan block \eqref{eq:a3}.
This means that $H^\times$ restricted to the subspace $M_{z_r}$ is a
Jordan operator of degree $r$ (in the standard notation it is rather the operator
$\frac1\Gamma H^\times$ which is called a Jordan operator).

These equations \eqref{eq:a3} are, like the eigenvector equation for
Dirac kets \eqref{eq:5e} and for Gamow vectors \eqref{eq:4.12} (Gamow
vectors of order 0 are Jordan vectors of degree 1), understood as
generalized eigenvector equations \eqref{eq:5f} over the space
$\Phi_+$, that means as
\begin{equation}\tag{A5} \label{eq:a5}
\begin{split}
\langle H\psi^- \lvert z_r^-\succ^{(k)} & \equiv 
\langle \psi^- \lvert H^\times\lvert z_r^-\succ^{(k)} \\
 & = z_r \langle \psi^- \lvert z_r^-\succ^{(k)} + \Gamma \langle
 \psi^- \lvert z_r^-\succ^{(k-1)} \quad \text{for all $\psi_- \in \Phi_+$.}
\end{split}
\end{equation}
Therefore the $\gket[r]{z_r^-}$ are generalized vectors in two
respects, firstly they are functionals on the space $\Phi_+$ and
secondly they are generalized eigenvectors as expressed by
\eqref{eq:a2}, \eqref{eq:a3}.
Therefore we call these vectors Gamow-Jordan vectors.
The matrix representation of the operator $H^\times$ is
given by a matrix that has in the diagonal complex eigenvalues for the ordinary Gamow kets
and Jordan blocks for the Gamow-Jordan kets.

For instance if the $S$-matrix has two first order
poles at $z=z_{R_k} =E_{R_k}-i\Gamma_{R_k}/2$, $k=1,2$, and one second
order pole at $z = z_2 = E_2 - i \Gamma_2$ then the matrix of the
Hamiltonian $H$ is given by
\begin{equation} \tag{A6}\label{eq:a6}
\begin{pmatrix} 
\langle H\psi^- \lvert z_2^-\succ^{(0)} \\
\langle H\psi^- \lvert z_2^-\succ^{(1)} \\
\ip{H\psi^-}{z_{R_1}} \\
\ip{H\psi^-}{z_{R_2}} \\
\ip{H\psi^-}{E}
\end{pmatrix} \equiv
\begin{pmatrix}
\langle \psi^- \lvert H^\times\lvert z_2^-\succ^{(0)} \\
\langle \psi^- \lvert H^\times\lvert z_2^-\succ^{(1)} \\
\ev{\psi^-}{H^\times}{z_{R_1}} \\
\ev{\psi^-}{H^\times}{z_{R_2}} \\
\ev{\psi^-}{H^\times}{E} \\
\end{pmatrix} =
\begin{pmatrix}
z_2    & 0    &         &         & \\
\Gamma_2      & z_2  &         &         & \\
       &      & z_{R_1} &         & \\
       &      &         & z_{R_2} & \\
       &      &         &         & (E)  
\end{pmatrix}\cdot
\begin{pmatrix}
\langle \psi^- \lvert z_2^-\succ^{(0)} \\
\langle \psi^- \lvert z_2^-\succ^{(1)} \\
\ip{\psi^-}{z_{R_1}} \\
\ip{\psi^-}{z_{R_2}} \\
\ip{\psi^-}{E}
\end{pmatrix}
\end{equation}
where $(E)$ denotes the continuously infinite matrix with diagonal
elements $E: E_0 \leq E < \infty$ corresponding to \eqref{eq:4.1}.
Each $z_{R_i}$ corresponds to \eqref{eq:4.12} and the $2 \times 2$
matrix in the top right corner is the Jordan block corresponding to
\eqref{eq:a3} with $r=2$.

The state associated to the pole term of the $S$-matrix for the $r$-th
order pole can no longer be described by a state vector, like the bound
states by $\ket{E_n}$ with real discrete eigenvalues, or the 1st
order resonance states (Gamow states) by $\ket{z_{R_i}^-}$ with complex
eigenvalue $z_{R_i}$ of the self-adjoint Hamiltonian $H$.
Instead, the state associated with the $r$-th order $S$-matrix pole term is
described by a non-diagonalizable density operator or state operator
\begin{equation}\tag{A7} \label{eq:a7}
W_{PT} = 2 \pi \Gamma \sum_{n=0}^{r-1} \binom{r}{n+1} (-i)^n W^{(n)}
\end{equation}
where the operators $W^{(n)}$ are defined as 
\begin{equation}\tag{A8} \label{eq:a8}
W^{(n)} = \sum_{k=0}^n \gket[k]{z_r^-} \gbra[n-k]{{}^-\!z_r}\ ,
\qquad n=0,1,2,\ldots, r-1\ .
\end{equation}
The pole term of the $r$th order $S$-matrix term is associated with a
sum \eqref{eq:a7} of the operators $W^{(n)}$.
The operators $W^{(n)}$
represent components of this state $W_{\text{PT}}$ which are in a
certain way, ``irreducible'' (as 
expressed by its property \eqref{eq:a19} below).

In the case $r=1$ (ordinary first order resonance pole) the operator
\eqref{eq:a7} becomes
\begin{equation}\tag{A9} \label{eq:a9}
W_{PT} = 2 \pi \Gamma \gket[0]{z_1^-}\gbra[0]{{}^-\!z_1} =
2 \pi \Gamma W^{(0)} = \ket{\psi^G}\bra{\psi^G}\ .
\end{equation}
This is the operator description of the generalized state whose vector
description is given by $\psi^G$ of \eqref{eq:4.12}.

For the case $r=2$ (second order pole at $z_R$) we have two
irreducible components:
\begin{equation} \tag{A10}\label{eq:a10}
W^{(0)} = \gket[0]{z_2^-}\gbra[0]{{}^-\!z_2} 
\end{equation}
and
\begin{equation}\tag{A11} \label{eq:a11}
W^{(1)} = \Big(\gket[0]{z_2^-}\gbra[1]{{}^-\!z_2}
+\gket[1]{z_2^-}\gbra[0]{{}^-\!z_2}  \Big)\ . 
\end{equation}
The state associated with the $n$th  order pole is a mixed state
$W_{PT}$ all of whose components $W^{(n)}$, except for the zeroth
component $W^{(0)}$ cannot be reduced further into something like
``pure'' states given by dyadic products like \eqref{eq:a9}   which
could be described by a vector $\psi^G=\gket[0]{z^-}\sqrt{2 \pi \Gamma}$.
The operator $W_{PT}$ associated to the 2nd order pole term is
\begin{equation}\tag{A12} \label{eq:a12}
\begin{split}
W_{PT} & = 2 \pi \Gamma \big(2 W^{(0)} - i W^{(1)}\big) \\
  & = 2 \pi \Gamma \Big(2 \gket[0]{z_2^-}\gbra[0]{{}^-\!z_2}
  \,-\, i \big(\gket[0]{z_2^-}\gbra[1]{{}^-\!z_2}
  \,+\,\gket[1]{z_2^-}\gbra[0]{{}^-\!z_2} \big) \Big) \ .
\end{split} 
\end{equation}

The generalized vectors $\gket[k]{z_r}$, $k=0,\ldots, r-1$ have
very complicated time evolution given by
\begin{equation}\tag{A13} \label{eq:a13}
e^{-iH^\times t}\gket[k]{z_r} = e^{-iz_r t} \sum_{\nu = 0}^k
\frac{\Gamma^\nu}{\nu!} (-it)^\nu \gket[k-\nu]{z_r}\ , \qquad t\geq 0\ .
\end{equation}

These are representations of the time transformation semigroup which
(for $r>1$) are not one dimensional.
The existence of this kind of representation for the causal spacetime
translation group has already been mentioned in reference \cite{ref27}.
For the special case of a double pole, $r=2$, $k=0,1$, the formula
\eqref{eq:a13} for the zeroth order Gamow vector is 
\begin{equation}\tag{A14} \label{eq:a14}
e^{-iH^\times t}\gket[0]{z_r^-} =e^{-iE_r t} e^{(-\Gamma/2)t}
\gket[0]{z_r^-}\ , \qquad t\geq 0\ ,
\end{equation}
and for the first order Gamow vector it is
\begin{equation}\tag{A15} \label{eq:a15}
e^{-iH^\times t}\gket[1]{z_r^-} = e^{-iz_r t} \Big(
\gket[1]{z_r^-} + (-it) \Gamma \gket[0]{z_r^-}\Big)\ , \qquad t\geq 0\ .
\end{equation}

That vectors associated with double poles of the $S$-matrix have in
addition to the exponential a strong linear dependence of magnitude
$\Gamma$ as in \eqref{eq:a15} or \eqref{eq:a13} has been known for a long time and was
the reason for dismissing double poles as viable resonance states, 
since the strong linear time dependence \eqref{eq:a15} of a deviation from the
exponential law had never been observed for decaying states.
However, since  the state associated with the $PT$ of the $S$-matrix
is not a vector state but given by the complicated density operator
\eqref{eq:a7}, \eqref{eq:a8} the relevant property is the time evolution of the state
operator $W_{PT}$ in \eqref{eq:a7}.
For the first order pole this is given according to \eqref{eq:a9} by
the operator equivalent of \eqref{eq:4.17} and \eqref{eq:4.18}.
Writing \eqref{eq:4.17} in terms of the state operator gives
\begin{equation}\tag{A16} \label{eq:a16}
\begin{split}
W^G(t) & = e^{-iH^\times t} \ket{\psi^G}\bra{\psi^G}e^{iHt} \\
 & = e^{-iz_Rt} \ket{\psi^G}\bra{\psi^G}e^{iz_R^*t} \\
 & = e^{-i(E_R-i(\Gamma/2))}\ket{\psi^G}\bra{\psi^G}
     e^{i(E_R+i(\Gamma/2))} \\
 & = e^{-\Gamma t}W^G(0)\ ,
\end{split} 
\end{equation}
The operator $W^G$ represents the microsystem that affects the
detector.
The vectors $\psi^-\in\Phi^+$ represent observables defined by the
detector (registration apparatus).
The probability that the microsystem affects the detector at $t>0$
(later than the time $t=0$ at which the microsystem was created) is
according to (2.1H)
\begin{equation}\tag{A17} \label{eq:a17}
\begin{split}
\Tr\big(\ket{\psi^-(t)}\bra{\psi^-(t)}W^G\big) & =
\ev{\psi^-(t)}{W^G}{\psi^-(t)} \\
  & = \ev{e^{-iHt}
    \psi^-}{W^G}{e^{iHt}\psi^-} \\
  & = \ev{\psi^-}{e^{-iH^\times t}W^Ge^{iHt}}{\psi^-} \\
  & = \ev{\psi^-}{W^G(t)}{\psi^-} \\
  & = e^{-\Gamma t} \ev{\psi^-}{W^G}{\psi^-}
\end{split} 
\end{equation}
which is \eqref{eq:4.18}.

We now apply the time evolution operator $e^{-iH^\times t}$ to the
state operator $W^{(n)}$ of \eqref{eq:a8}, $n=0,1,\ldots, (r-1)$ and
then the $W_{PT}$ of \eqref{eq:a7}, which is the state operator
associated to the $S$ matrix pole term of order $r$ (any finite
integer):
\begin{equation}\tag{A18} \label{eq:a18}
\begin{split}
W^{(n)}(t) & = e^{-iH^\times t} W^{(n)} e^{iHt} \\
  & = \sum_{k=0}^n e^{-iH^\times t} \gket[k]{z_r^-}
  \gbra[k-n]{{}^- z_r} e^{iHt}\ .
\end{split} 
\end{equation}
Using \eqref{eq:a13} and its conjugate
\begin{equation} 
\tag{$\overline{A13}$} \label{eq:a13b}
\gbra[k]{{}^- z_r}e^{iH t} = e^{iz_r^* t} \sum_{\nu = 0}^k
\frac{\Gamma^\nu}{\nu!} (it)^\nu\  \gbra[k-\nu]{{}^- z_r}
\end{equation}
one obtains after a complicated calculation \cite{12} a very simple expression
\begin{equation}\tag{A19} \label{eq:a19}
W^{(n)}(t) = e^{-iH^\times t}W^{(n)}e^{iHt} = e^{-\Gamma t} \sum_{k=0}^n
\gket[k]{z^-}\gbra[k-n]{{}^- z } = e^{-\Gamma t} W^{(n)}(0)\ ,
\quad t\geq 0\ .
\end{equation}
The complicated state operator $W^{(n)}$ has thus a very simple exponential
time evolution. 
Considering the complicated time evolution of
\eqref{eq:a13} and \eqref{eq:a13b} the simple result \eqref{eq:a19} looks like
a miracle.

This result means that the complicated non-reducible (i.e., ``mixed'')
microphysical state operator $W^{(n)}$ defined by \eqref{eq:a8} has a
simple purely exponential semigroup time evolution, like the zeroth
order Gamow state \eqref{eq:a16} and thus leads to the exponential law
for the probabilities, as in \eqref{eq:a17}.
This operator is probably the only operator formed by the dyadic
products $\gket[m]{z_r^-} \gbra[\ell]{{}^-z_r}$ with $m,\ell =
0,1,\ldots, n$, which has purely exponential time evolution.
Thus $W^{(n)}$ of equation \eqref{eq:a8} is distinguished from all
other operators in $\cM^{(n)}_{z_r}$.

The microphysical decaying state operator associated with the $r$-th
order pole of the unitary $S$-matrix is according to its definition
\eqref{eq:a7} a sum of the $W^{(n)}$.
Because of the simple form \eqref{eq:a19} (independence of the time
evolution of $n$) this sum has again a simple and exponential time
evolution 
\begin{equation}\tag{A20} \label{eq:a20}
W_{PT}(t) \equiv e^{-iH^\times t} W_{PT}e^{iHt} = e^{-\Gamma t}
W_{PT}\ ; \qquad t\geq 0\ .
\end{equation}
Thus we have seen that the state operator \eqref{eq:a7} which is the
operator associated to
the $r$-th order pole of the $S$-matrix,  describes a non-reducible ``mixed''
microphysical decaying state which obeys an exact exponential decay
law.

Summarizing the Appendix,
Jordan blocks arise naturally from higher order $S$-matrix poles.
They represent a self-adjoint Hamiltonian by a
complex matrix in a finite dimensional subspace like the two
dimensional Jordan block in \eqref{eq:a6}. 
This finite dimensional subspace  is contained in the dual
$\Phi_+^\times$ of the
Rigged Hilbert space of Hardy type \eqref{eq:12}.
Although higher order $S$-matrix poles are not excluded by any theoretical
argument, there has been so far very little experimental evidence for
their existence.
It was always believed on the basis of \eqref{eq:a15} that 
states associated with higher order poles must
have polynomial time dependence and therewith deviations from the exponential
time dependence.
A deviation from the exponential law of the magnitude as predicted by
\eqref{eq:a15} (i.e., of magnitude $\Gamma$) is excluded experimentally.
However, since \eqref{eq:a19} and \eqref{eq:a20} show that all non-reducible
state operators associated to the higher order $S$-matrix pole evolve
purely exponential in time, there remains no
experimental evidence against their existence.
These mathematically beautiful objects may therefore have some
future applications in physics.

\end{appendix}

\end{document}